\documentclass[twocolumn]{aastex631}

\usepackage{array, enumitem, natbib}

\begin{document}

\title{Pulling back the curtain on shocks and star-formation in NGC\,1266 with Gemini-NIFS}

\author[0000-0003-3191-9039]{Justin Atsushi Otter}
\email{jotter2@jhu.edu}
\affiliation{William H. Miller III Department of Physics and Astronomy, Johns Hopkins University, Baltimore, MD 21218, USA}

\author[0000-0002-4261-2326]{Katherine Alatalo}
\affiliation{Space Telescope Science Institute, 3700 San Martin Drive, Baltimore, MD 21218, USA}
\affiliation{William H. Miller III Department of Physics and Astronomy, Johns Hopkins University, Baltimore, MD 21218, USA}

\author[0000-0001-7883-8434]{Kate Rowlands}
\affiliation{AURA for ESA, Space Telescope Science Institute, 3700 San Martin Drive, Baltimore, MD 21218, USA}
\affiliation{William H. Miller III Department of Physics and Astronomy, Johns Hopkins University, Baltimore, MD 21218, USA}

\author[0000-0002-8175-7229]{Richard M. McDermid}
\affiliation{Astrophysics and Space Technologies Research Centre, School of Mathematical and Physical Sciences, Macquarie University, Sydney, Australia}
\affiliation{ARC Centre of Excellence for All Sky Astrophysics in 3 Dimensions (ASTRO 3D)}

\author[0000-0003-4932-9379]{Timothy A. Davis}
\affiliation{Cardiff Hub for Astrophysics Research \& Technology, School of Physics \& Astronomy, Cardiff University, Queens Buildings, Cardiff CF24 3AA, UK}

\author[0000-0002-0706-2306]{Christoph Federrath}
\affiliation{Research School of Astronomy and Astrophysics, Australian National University, Canberra, ACT 2611, Australia}
\affiliation{ARC Centre of Excellence for All Sky Astrophysics in 3 Dimensions (ASTRO 3D)}

\author[0000-0002-4235-7337]{K. Decker French}
\affiliation{Department of Astronomy, University of Illinois, 1002 W. Green Street, Urbana, IL 61801, USA}

\author[0000-0001-6670-6370]{Timothy Heckman}
\affiliation{William H. Miller III Department of Physics and Astronomy, Johns Hopkins University, Baltimore, MD 21218, USA}
\affiliation{School of Earth and Space Exploration, Arizona State University, }

\author[0000-0002-3471-981X]{Patrick Ogle}
\affiliation{Space Telescope Science Institute, 3700 San Martin Drive, Baltimore, MD 21218, USA}

\author[0000-0002-2603-2639]{Darshan Kakkad}
\affiliation{Space Telescope Science Institute, 3700 San Martin Drive, Baltimore, MD 21218, USA}

\author[0000-0002-0696-6952]{Yuanze Luo}
\affiliation{William H. Miller III Department of Physics and Astronomy, Johns Hopkins University, Baltimore, MD 21218, USA}

\author[0000-0003-1991-370X]{Kristina Nyland}
\affiliation{U.S. Naval Research Laboratory, 4555 Overlook Ave SW, Washington, DC 20375, USA}

\author[0000-0002-6582-4946]{Akshat Tripathi}
\affiliation{Department of Astronomy, University of Illinois, 1002 W. Green Street, Urbana, IL 61801, USA}

\author[0000-0002-9471-8499]{Pallavi Patil}
\affiliation{William H. Miller III Department of Physics and Astronomy, Johns Hopkins University, Baltimore, MD 21218, USA}

\author[0000-0003-4030-3455]{Andreea Petric}
\affiliation{Space Telescope Science Institute, 3700 San Martin Drive, Baltimore, MD 21218, USA}

\author[0000-0003-2599-7524]{Adam Smercina}
\affiliation{Astronomy Department, University of Washington, Seattle, WA 98195, USA}

\author[0009-0004-0844-0657]{Maya Skarbinski}
\affiliation{William H. Miller III Department of Physics and Astronomy, Johns Hopkins University, Baltimore, MD 21218, USA}

\author[0000-0002-3249-8224]{Lauranne Lanz}
\affiliation{Department of Physics, The College of New Jersey, Ewing, NJ 08628, USA}

\author[0000-0003-3917-6460]{Kristin Larson}
\affiliation{Space Telescope Science Institute, 3700 San Martin Drive, Baltimore, MD 21218, USA}

\author[0000-0002-7607-8766]{Philip N. Appleton}
\affiliation{Caltech/IPAC, MC 314-6, 1200 E. California Blvd., Pasadena, CA 91125, USA}

\author{Susanne Aalto}
\affiliation{Department of Space, Earth and Environment, Chalmers University of Technology, 412 96 Gothenburg, Sweden}

\author[0000-0003-4830-9069]{Gustav Olander}
\affiliation{Department of Space, Earth and Environment, Chalmers University of Technology, 412 96 Gothenburg, Sweden}

\author[0000-0001-6245-5121]{Elizaveta Sazonova}
\affiliation{Department of Physics and Astronomy, University of Waterloo, 200 University Avenue West, Waterloo, ON N2L 3G1, Canada}

\author[0000-0003-1545-5078]{J. D. T. Smith}
\affiliation{University of Toledo Department of Physics and Astronomy, Ritter Astrophysical Research Center, Toledo, OH 43606, USA}

\begin{abstract}
We present Gemini near-infrared integral field spectrograph (NIFS) K-band observations of the central 400\,pc of NGC\,1266, a nearby (D$\approx$30 Mpc) post-starburst galaxy with a powerful multi-phase outflow and a shocked ISM.
We detect 7 H$_2$ ro-vibrational emission lines excited thermally to $T$$\sim$2000 K, and weak Br$\gamma$ emission, consistent with a fast C-shock. 
With these bright H$_2$ lines, we observe the spatial structure of the shock with an unambiguous tracer for the first time.
The Br$\gamma$ emission is concentrated in the central $\lesssim$100\,pc, indicating that any remaining star-formation in NGC\,1266 is in the nucleus while the surrounding cold molecular gas has little on-going star-formation.
Though it is unclear what fraction of this Br$\gamma$ emission is from star-formation or the AGN, assuming it is entirely due to star-formation we measure an instantaneous star-formation rate of 0.7 M$_\odot$ yr$^{-1}$, though the star-formation rate may be significantly higher in the presence of additional extinction.
NGC\,1266 provides a unique laboratory to study the complex interactions between AGN, outflows, shocks, and star-formation, all of which are necessary to unravel the evolution of the post-starburst phase.
\end{abstract}


\section{Introduction} \label{sec:intro}

The relationship between a galaxy and its supermassive black hole holds many of the keys to unlocking how galaxies evolve over cosmic time.
Accreting supermassive black holes release enormous amounts of energy in the form of radiation pressure driven winds from the accretion disk or via collimated radio jets \citep[e.g.][]{fabian12,heckman14, heckman23}.
The connection between these actively accreting supermassive black holes and star-formation in their host galaxies is often referred to as active galactic nuclei (AGN) feedback.
Though feedback can be positive (instigating star-formation), negative feedback (suppressing star-formation) is necessary in theoretical work to regulate star-formation and to reproduce the galaxy stellar mass function and the low overall star-formation efficiency in the Universe \citep[e.g.][]{kauffmann99, dimatteo05, hopkins06a, vogelsberger14, schaye15}.
Direct observational evidence of AGN feedback has mostly centered on massive elliptical galaxies, and while indirect evidence exists in lower mass galaxies, direct evidence is scant and results are mixed \citep[e.g.][]{martin07, schawinski10, wild10, fabian12, cheung16, roy18, smethurst18, french18a}.
Further, much work remains to build a precise physical understanding of the connections and interplay between the black hole accretion, the interstellar medium (ISM), and star-formation in galaxies.

Outflows are one of the clearest mechanisms by which an AGN interacts with the host galaxy on kpc scales though their impact on the host galaxy's ISM and star-formation is unclear.
Substantial observational and theoretical evidence demonstrates that large scale outflows and jets can slowly quench the host galaxy by heating the surrounding circumgalactic medium, thus preventing further cold gas accretion and ending star-formation once the gas reservoirs are depleted \citep[e.g.][]{birzan04, gaspari11, morganti13, weinberger17}.
However, the impacts of smaller kpc-scale outflows on the host galaxy's ISM remain uncertain. 
While some outflows have sufficient velocity and mass to effectively remove the gas reservoirs from the host galaxy \citep{nesvadba07, harrison14, rupke17}, other outflows will not escape the host galaxy \citep{alatalo15, fernandez-ontiveros20, luo22}.
Still, these non-ejective outflows can have important impacts on the galaxy's star-formation properties as they can heat and drive turbulence into the ISM \citep{nesvadba10, nesvadba11a, guillard12a, mandal21a}, though it is not apparent whether this heating can impact star-formation across the entire galaxy \citep{gabor14, bae17}

Theoretical and observational work has shown that the turbulence and shock properties in molecular gas are key in determining the star-formation efficiency of that gas. 
Supersonic turbulence and resulting shocks create compressions in the gas, widening the distribution of gas densities.
While the high density gas created by these compressions is ripe for star-formation, turbulent stirring motions counteract star-formation by hindering gravitational collapse \citep{federrath12}.
The relationship between turbulence and star-formation is primarily driven by the virial parameter of the gas (two times the ratio of turbulent to gravitational energy), the sonic Mach number (the ratio of velocity dispersion to sound speed), and the driving mode of the turbulence (the ratio of the compressive and solenoidal, or stirring, modes).

Both galactic and extragalactic observational studies demonstrate that physical processes that impact these turbulence parameters can effectively suppress star-formation.
Molecular clouds in the central molecular zone of the Milky Way have low star-formation rates for the molecular gas mass, which may be caused by solenoidal mode turbulence driven by shear forces \citep{federrath16, barnes17}.
In an extragalactic context, \citet{salim20} show that the observed star-formation suppression in NGC 7674 is consistent with an escalated virial parameter and solenoidal mode turbulence driven by shear forces from the stellar bar and dense group environment.
High turbulent pressures have been observed in other post-starburst galaxies, though further observations are necessary to directly link this turbulence to the ongoing star-formation suppression \citep{smercina22}.

The relationships between AGN, radio jets, outflows, and the host galaxy's ISM and star-formation form a complex, interconnected web.
While progress has been made in studying individual connections in this web, an integrated understanding of the entire system is lacking.
NGC\,1266 provides an ideal laboratory for beginning to unpack these relationships in detail.
NGC\,1266 is a nearby ($D$=29.9 Mpc, 1\arcsec\,= 0.14 kpc) post-starburst galaxy with a powerful multi-phase outflow and a sizable molecular gas reservoir. 
Though NGC\,1266 visually appears to be a typical S0 galaxy, CO observations reveal a large, centrally concentrated molecular gas reservoir ($1.7\times10^{9}$ M$_\odot$) and a molecular outflow, achieving a molecular outflow mass rate of 110 M$_\odot$/yr \citep{alatalo11, alatalo15}.
Despite high molecular gas surface densities and the presence of dense gas in the central region, there is little ongoing star-formation, as the total star-formation rate is suppressed by a factor of 10-100 times from what is expected of the molecular gas mass using the Kennicutt-Schmidt relation \citep{alatalo15}.
The outflow is multi-phase including ionized, atomic, molecular, and dense molecular gas components \citep{davis12}.
The orientation of the outflow is similar to extended radio emission, which may be tracing relativistic radio jets from the AGN \citep{nyland13}.
It is likely that an obscured, low-luminosity AGN drives this outflow, potentially through radio jets, as the outflow mass rate far exceeds even the most generous star-formation rate measurements, and the radio jets have sufficient power to drive the outflow \citep{nyland13, alatalo15}.

The impact of the outflow on the ISM of NGC\,1266 remains a critical open question necessary to understand the role of AGN feedback in this galaxy.
Both the ionized gas emission lines and highly excited molecular transitions (CO, H$_2$O) demonstrate that the ISM is dominated by shocks \citep{davis12, pellegrini13}.
While the powerful outflow is an intuitive driver of these shocks, the lack of spatially resolved observations of the shock features prevents definitive statements on the nature of the interactions between the outflow and ISM.
Further, previous studies suggest that outflow-driven turbulence is suppressing star-formation in the molecular gas, though high spatial resolution observations of the outflow and shocks are necessary to verify this scenario \citep{alatalo15}.

In NGC\,1266, we aim to uncover the relationship between the outflow, ISM, and star-formation and thus the physics underlying AGN feedback.
Spatially resolved NIR integral field spectroscopy (IFS) observations provide a crucial path toward this goal, as these data provide important constraints on the shock, molecular gas, and star-formation properties of the ISM.
Bright NIR ro-vibrational H$_2$ emission lines trace warm (T$\gtrsim$1000K) molecular gas, which can be heated by star formation, around HII regions, and by non-star forming processes, such as shocks.
Thermal excitation sources for H$_2$ include shocks and UV or X-ray radiation, which collisionally excite the gas, whereas UV or X-ray fluorescence are the primary non-thermal excitation mechanisms \citep[e.g.][]{kwan77, lepp83, sternberg89, draine90, dors12, u19a}.
Non-thermal excitation mechanisms may still yield thermalized $\nu$=1 states (i.e. (1-0) ro-vibrational emission lines), though the $\nu=2$ states (i.e. (2-1) emission) will be more populated than would be expected from the $\nu$=1 column densities. 

As differing star-formation tracers yield a wide range of star-formation rates in NGC\,1266, NIR Hydrogen recombination lines provide a unique star-formation constraint.
This variance of measured star-formation rates likely arises from differing timescales probed by different methods.
While some methods, such as Hydrogen recombination lines, measure star-formation on $<10$ Myr timescales, other methods, such as the far-IR luminosity, measure the integrated star-formation rate over a much longer period of time, which may include the recent burst of star-formation in a post-starburst galaxy \citep{french21}. 
NIR Hydrogen recombination lines, such as Br$\gamma$, provide a reliable, instantaneous star-formation tracer less impeded by dust than their optical counterparts.
The combination of Br$\gamma$ and H$_2$ rovibrational lines further allows us to determine whether the Br$\gamma$ we observe is dominated by shocks. 

In this work, we present adaptive optics enhanced K-band Gemini near-infrared integral field spectrograph (NIFS) observations of the central 400\,pc of NGC\,1266, covering multiple H$_2$ ro-vibrational transitions as well as the hydrogen recombination line Br$\gamma$. 
In Section~\ref{sec:obs}, we provide details of these observations and data reduction.
In Section~\ref{sec:results}, we present our emission line measurements and plot various line diagnostics.
We discuss possible H$_2$ excitation mechanisms, shock properties and structure, and the nuclear properties in Section~\ref{sec:discuss}, and share our conclusions in Section~\ref{sec:conclusion}.
When computing distances and scales, we adopt a Hubble constant of $H_0$=70km s$^{-1}$ Mpc$^{-1}$, $\Omega_m = 0.3$, and $\Omega_\Lambda = 0.7$.

\section{Observations and Methods} \label{sec:obs}

\subsection{Gemini-NIFS Observations and Reduction}

Near-infrared observations of NGC\,1266 were obtained with NIFS with laser guide star adaptive optics enabled on the Gemini North Telescope \citep{mcgregor03, christou10} under program ID GN-2013B-Q-30 (P.I. McDermid).
These observations were taken with the K-band grating (central wavelength $\lambda$ = 2.2 $\mu$m) in three observing sessions from November 2014 to February 2015.
The NIFS field of view is 3\arcsec\,$\times$\,3\arcsec\, with an angular sampling of 0\arcsec.103 $\times$ 0\arcsec.043.
The final science cubes are rebinned to a square pixel size of 0.043\arcsec.
The spectral resolution of NIFS in the K-band is R$\sim$5290.
With adaptive optics, NIFS observations typically achieve spatial resolutions of 0.1\arcsec-0.2\arcsec. 

Our observing strategy was as follows.
In each observing session, an A-type star was observed immediately before NGC\,1266 at a comparable airmass to facilitate the correction of telluric absorption.
Each science exposure was 600\,s, with a total of 24 exposures split across the three nights, for a total exposure time of 4 hours. 
Science exposures in each observing session were dithered to minimize the impact of bad detector pixels. 
Immediately after science observations, Ar and Xe arc lamp calibration exposures were taken for wavelength calibration.
In the latter two observing sessions, a different A-type star was additionally observed immediately after science observations.
Other standard calibration exposures were taken including flat fields, darks, and a Ronchi calibration mask exposure to account for spatial distortion.
More details on each observing session are provided in Table~\ref{tab:obs}.

The data were reduced with \texttt{Nifty4Gemini}, a python pipeline built upon Gemini IRAF \citep{lemoine-busserolle19}.
The primary reduction steps for the science observations include co-adding the science frames, performing sky subtraction, identifying and interpolating over bad pixels, computing spatial and wavelength distortion maps, correcting for telluric lines, flux calibrating the cube, and finally merging the resulting science data from each observing visit.
The wavelength solution is determined from the arc lamp exposures. 
Telluric corrections are performed with the A-type star observations.
The flux is calibrated by fitting a blackbody function to the observed A-type star spectrum at a known temperature, and then scaling the blackbody function to the previous measured 2MASS magnitude of the star.
The flux calibration of NIFS is accurate to approximately 10\%, as stated in the online documentation\footnote{\url{https://www.gemini.edu/observing/resources/near-ir-resources}}.

The automatically generated offset values for merging data cubes in the pipeline were not accurate, so we specified integer pixel offsets manually. 
We determined these offsets by comparing the isophotes of continuum emission in each cube. 
The merging was performed with the pipeline, which uses the \textit{imcombine} IRAF task to combine the cubes.
This task only accepts integer pixel offsets, so we could not merge the cubes with sub-pixel accuracy.
We had to merge both the cubes for different dither positions in each observing session, as well as the resulting three cubes from each observing session into one final science-ready data cube.
Due to the manual offset specification, the final merged cube lacked absolute astrometry information.
We register the NIFS data to accompanying F160W (H-band) \textit{Hubble Space Telescope} (\textit{HST}) from \citet{alatalo14} by matching isophotes of the continuum emission.
We assume the peak of the H-band continuum from \textit{HST} is coincident with the NIFS K-band continuum, which is likely given they are both in the NIR.

Lastly, we find sharp absorption features in specific channels due to poor sky subtraction on each observing night. 
None of these features fall on the peak of detected emission lines, so we mask these channels in our analysis and do not expect them to have a significant impact on our results. 

\begin{table*}
    \centering
    \begin{tabular}{c|c|c|c|c|c|c}
    Date & Exposure  & Exposures & Total  & Number of  & Uncorrected & Calibration  \\ 
     &  length &  &  Exposure time & dithers &  V-band seeing &  standard stars \\ 
    \hline
    Nov. 10 2014 & 600s & 7 & 4200s & 4 & 0.33\arcsec & HIP10512  \\ \hline
    Jan. 31 2015 & 600s & 6 & 3600s & 4 & 0.49\arcsec & HIP10512, HIP22923  \\ \hline
    Feb. 1 2015 & 600s & 11 & 6600s & 6 & 0.69\arcsec & HIP10512, HIP22923 
    \end{tabular}
    \caption{Details of each observing session with NIFS.}
    \label{tab:obs}
\end{table*}

\subsection{Spectral fitting} \label{sec:fitting}

To accurately model the stellar continuum, we use the \texttt{pPXF} python package, an implementation of the  penalized pixel-fitting method of \citet{cappellari23}, building upon previous versions of the algorithm \citep{cappellari04, cappellari17}. 
We use the X-shooter spectral library (XSL) data release 3 as stellar templates \citep{verro22}.
This library consists of 830 high resolution ($R\sim$10,000) stellar spectra of 683 stars.
These spectra span UV to NIR wavelengths, from 0.35 - 2.480 $\mu$m, fully encompassing the wavelength range of our NIFS observations.
We exclude XSL spectra which have not been corrected for wavelength-dependent slit flux loss, and spectra which were not corrected for galactic extinction, resulting in 670 remaining spectra.
Stars in XSL were chosen to span as much of the Hertzsprung-Russel diagram as possible with a wide range of chemical compositions.
Cool and otherwise infrared-luminous stars are emphasized, making XSL an ideal choice for NIR stellar continuum fitting.
When fitting the stellar continuum, we mask regions around H$_2$ emission lines and Br$\gamma$.
To model the observed stellar continuum as closely as possible, we include an additive and multiplicative Legendre polynomial of degree 10 to account for template mismatch.

At the ends of the wavelength range, above 2.42 $\mu$m and below 2.025 $\mu$m, the fitted continuum sharply increases due to the multiplicative Legendre polynomial.
Because this upturn in the stellar continuum heavily impacts our measured flux of the H$_2$(1-0) Q(1) line, we instead use the median of the spectrum from 2.405 - 2.419 $\mu$m as a flat continuum value.
This wavelength range is adjacent to the H$_2$(1-0) Q(1) line and is free from major emission and absorption features.
We note that the measured fluxes of the H$_2$(1-0) Q(1) line are less reliable due to this less precise continuum fitting method.
We similarly replace the continuum below 2.025\,$\mu$m with the median value in the line-free region of 2.025 - 2.043\,$\mu$m.

To measure emission line properties, we fit emission lines using the following procedure.
We perform a one component and two component Gaussian fit on the continuum subtracted spectrum.
We use the trust region reflective fitting algorithm, a non-linear least squares fitting method, as implemented by \texttt{scipy}.
To determine whether a second Gaussian component represents a statistical improvement over the single component fit, we compute the Bayesian Information Criterion, assuming Gaussian errors as:
\begin{equation}
    BIC = n \ln \left(\frac{\text{SSR}}{n}\right) + k \ln(n)
\end{equation}
where $n$ is the number of observations, SSR is the sum of squared residuals between the model and observations, and $k$ is the degrees of freedom in the model.
To accept the two component model, we require 
\begin{equation}
    BIC_{1} - BIC_{2} > 10
\end{equation}
where BIC$_{1}$ and BIC$_2$ correspond to the one- and two-component models, respectively.
The BIC for each model is not informative, only the difference between the two values provides a heuristic for which model better suits the data.
We use a threshold value of 10, indicating there is very strong evidence that a second Gaussian component is needed to adequately represent the data \citep[e.g.][]{mukherjee98}.
From visual inspection, this threshold value matches by-eye determinations of whether one or two components are needed.

We compute uncertainties for our fluxes by measuring the standard deviation of a line-free continuum subtracted region of the spectrum from 2.26 - 2.35 $\mu$m.
We use the same spectral region for all of our uncertainties because some of our lines are in spectral regions with closely-spaced emission lines, making it difficult to reliably measure the standard deviation of these regions.
We convolve this uncertainty with the aforementioned 10\% flux calibration uncertainty in our analysis.

To measure upper limit fluxes for non-detected emission lines, we assume a linewidth and an amplitude of three times the standard deviation of a continuum-subtracted line-free region of the spectrum.
For H$_2$ lines, we assume the same linewidth as the narrow component of H$_2$(1-0) S(1), the brightest H$_2$ line.
For He I and Br$\gamma$, we adopt a linewidth of 130 km s$^{-1}$, based roughly on the average linewidth of the narrow component of optical ionized gas emission lines from \citet{davis12}.

We find that there is a sky over-subtraction feature near Br$\gamma$.
To ensure that this feature does not impact our measured Br$\gamma$ fluxes, we determine the channels where this feature is present by comparing spectra across the cube, including in regions without any Br$\gamma$ emission.
Because this feature is only on the edge of the Br$\gamma$ line profile, we mask these contaminated channels and fit Br$\gamma$ without severely impacting the Gaussian fit and resulting flux measurement.

To fit all spaxels in our data cube, we traverse the cube in a spiral starting from the center, and use the previous fit values as starting values for fitting the next spaxel.

\section{Results} \label{sec:results}

\subsection{Detected lines}
\begin{figure*}
    \centering
    \includegraphics[width=\textwidth]{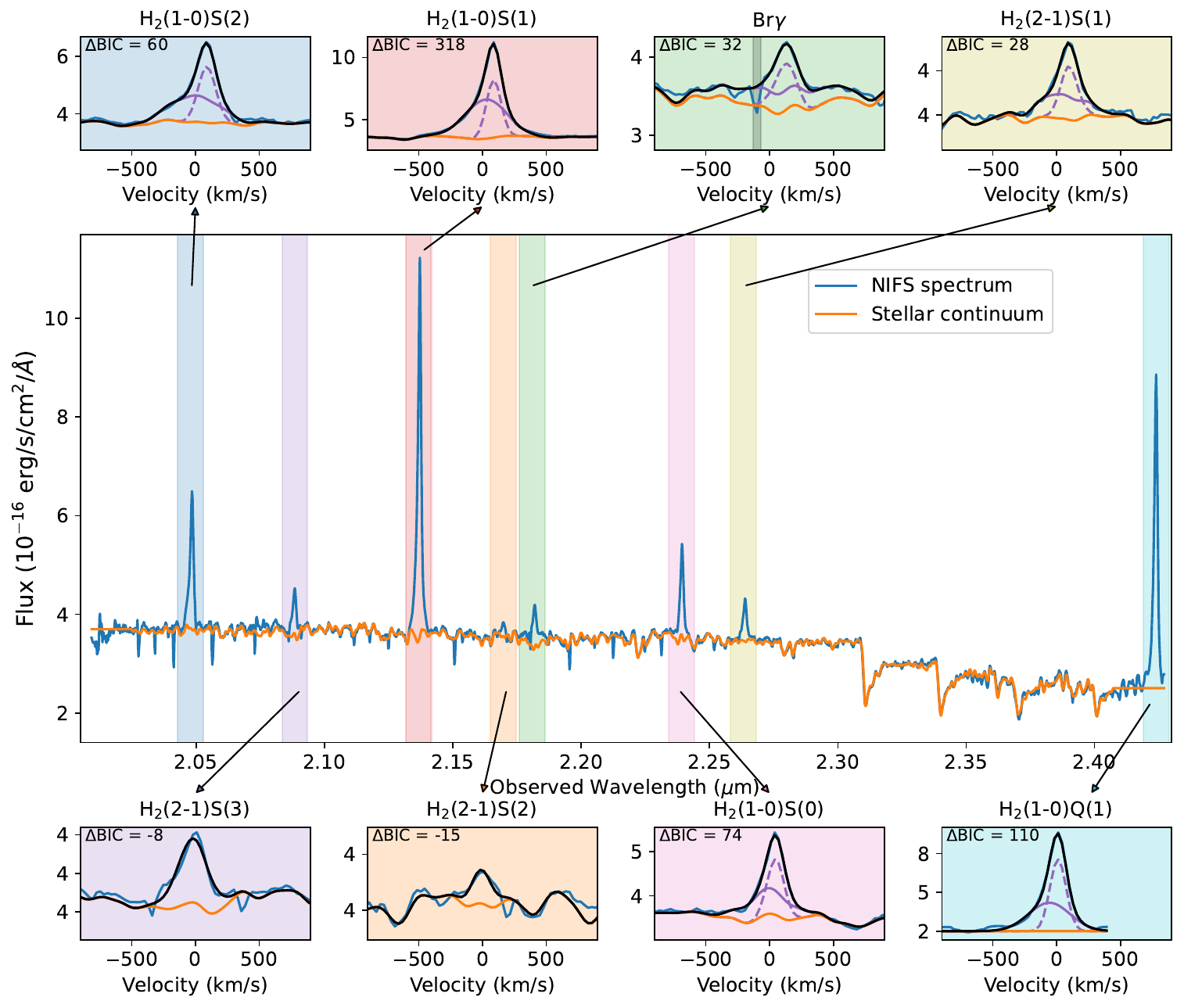}
    \caption{K-band spectrum of the central 100\,pc of NGC\,1266 from Gemini-NIFS. The surrounding axes show zoom-ins of detected emission lines where the colors of the axes show the corresponding location in the spectrum. The purple lines show each fitted Gaussian component, and the orange line is the linear continuum fit. The black line is the total two-component Gaussian and continuum fit. In the Br$\gamma$ panel, the gray shaded region shows points masked out due to sky over-subtraction.}
    \label{fig:cent_spec}
\end{figure*}

We detect 7 ro-vibrational H$_2$ emission lines and Br$\gamma$ in the nucleus of NGC\,1266. 
Figure~\ref{fig:cent_spec} shows the K-band NIFS spectrum of NGC\,1266 extracted within a circular aperture at the center of the galaxy with a diameter of 100\,pc (0.7\arcsec). 
Highlighted regions show the 8 detected emission lines, with an inset for each, showing a two-component Gaussian fit to the emission line.
Two Gaussian components are required to fit the line profiles as shown in Figure~\ref{fig:cent_spec} where a wide, blue-shifted component is detected in the brightest H$_2$ lines.

\begin{figure*}[!h]
    \centering
    \includegraphics[width=\textwidth]{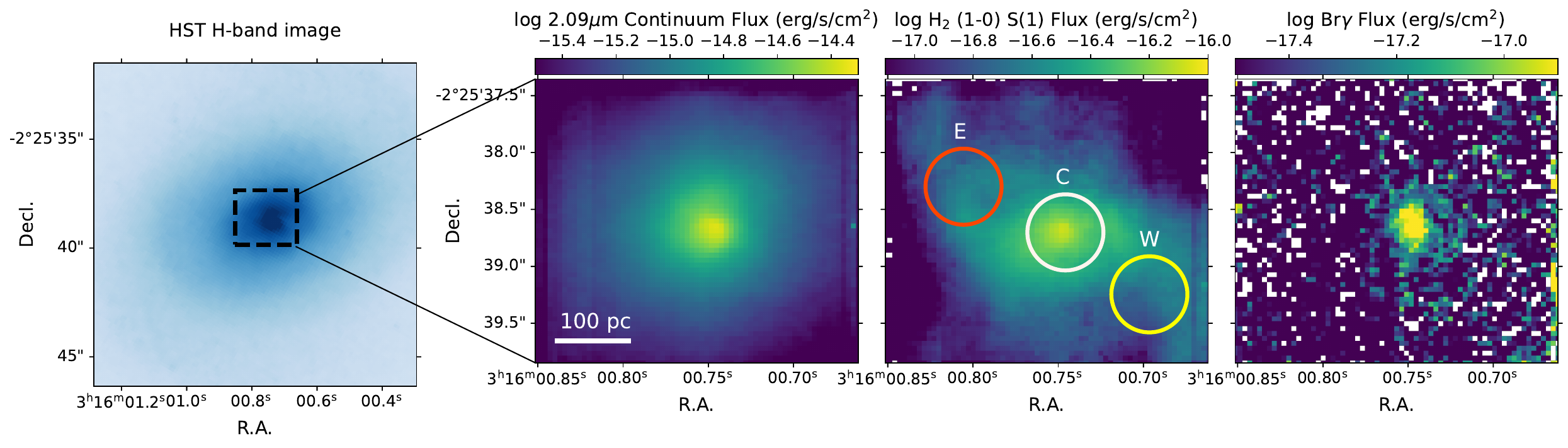}
    \caption{Left: \textit{HST} H-band image of NGC\,1266. The black dashed rectangle shows the Gemini-NIFS field of view. Center Left: NIFS log 2.09$\mu$m continuum flux map. The continuum is measured in a line-free spectral region. Center Right: log H$_2$ (1-0) S(1) flux map. The three circles show three apertures extracted. Each aperture has a physical radius of 50\,pc, and is labelled W, E, or C. Line fluxes are measured by summing the two fitted Gaussian components and are continuum subtracted. Right: Br$\gamma$ flux map. }
    \label{fig:flux_maps}
\end{figure*}

Selected results of the full data cube Gaussian fitting are shown in Figure~\ref{fig:flux_maps}.
The left panel is an integrated flux map of the line-free region at 2.09\,$\mu$m.
The middle panel shows the fitted H$_2$(1-0) S(1) flux map, the brightest H$_2$ emission line.
The three circles in this panel are three different apertures used in later analysis to show the emission line properties in different regions: east, west, and center.
These apertures are chosen to sample different spatial regions of the shock and outflow: the nucleus, and the two extended wings of H$_2$ emission.
Finally, the third panel shows the fitted Br$\gamma$ flux map.
Spaxels with Br$\gamma$ S/N $<$ 3 are not shown.
A thin line on the west side visible in all three images is a detector artifact, so we avoid this region in the west aperture and neglect the bright Br$\gamma$ emission on the edge of the detector.

We measure fluxes in each aperture by summing the spectra and fitting two Gaussian components as described in Section~\ref{sec:fitting}. 
Table~\ref{tab:flux_table} shows the measured fluxes for each emission line in each aperture and across the NIFS field of view. 
We do not detect He I 2.06 $\mu$m emission line in any of our apertures or summing across the entire field of view, so we report measured upper limits.
Flux uncertainties reported in this table are from fitting only and do not include flux calibration uncertainties.
We repeat our analysis with non-parametric fluxes measured by summing the continuum subtracted spectra within $\pm$800 km/s of each line, and find similar, though noisier, results.

We repeat our analysis with apertures chosen at different locations across the NIFS field of view and find little variation in our results.
Therefore, we have only plotted the three apertures shown in Figure~\ref{fig:flux_maps} as these apertures are representative and they have a high S/N.

\subsection{$H_2$ Kinematics} \label{sec:kinematics}

\begin{figure*}
    \centering
    \includegraphics[width=0.85\textwidth]{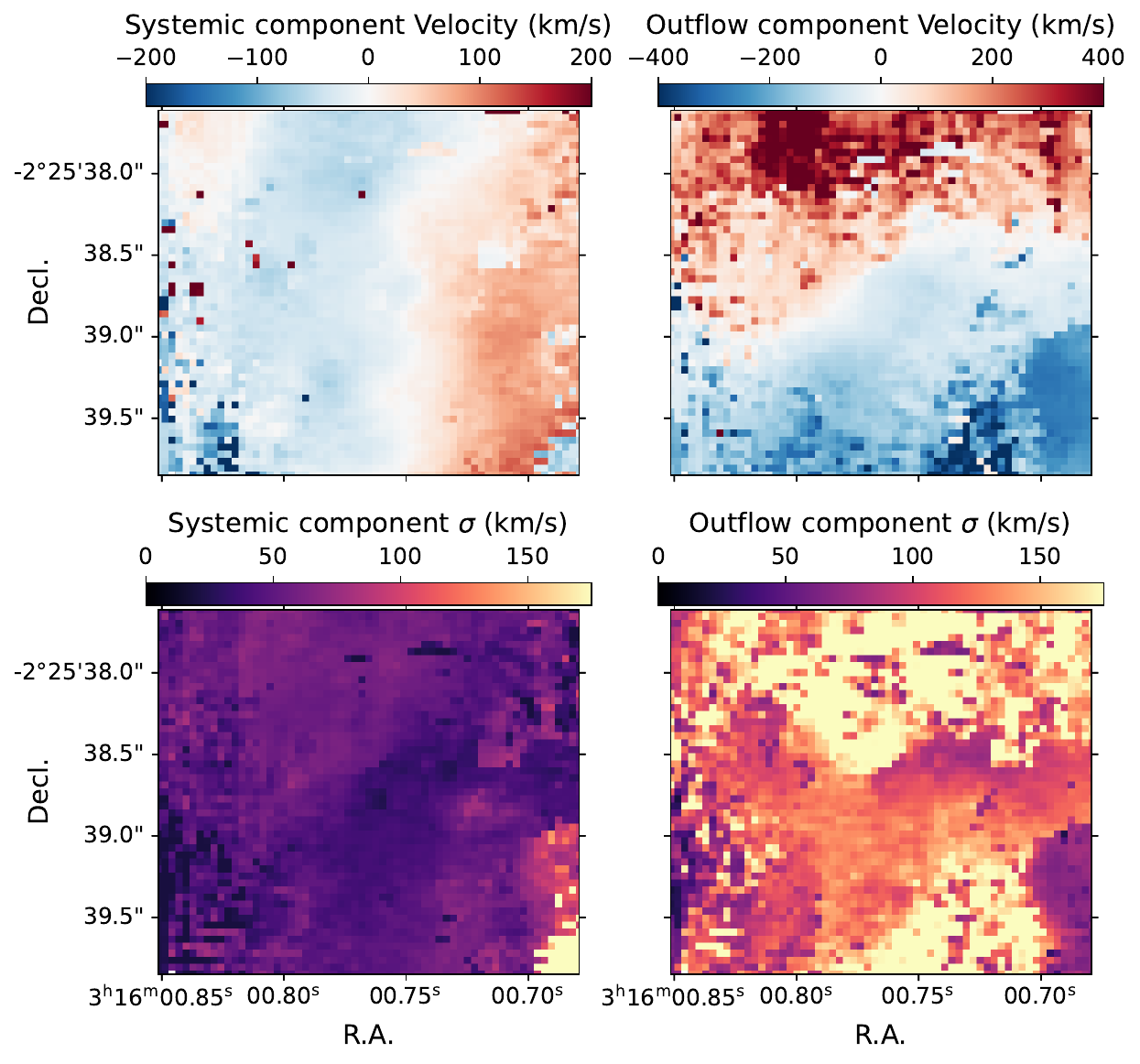}
    \caption{Fitted two-component kinematics of H$_2$(1-0)S(1). The top row shows the velocity maps and the lower row the measured linewidth. The velocity field of the first component on the left is similar to the stellar velocity field thus tracing the systemic gas rotation, while the second, broad component on the right traces the outflow and is misaligned from the systemic component.}
    \label{fig:velmaps}
\end{figure*}

We plot our measured velocities and linewidths from our two-component Gaussian fitting described in Section~\ref{sec:fitting} in Figure~\ref{fig:velmaps}.
We begin by setting the narrower Gaussian component as component 1, and broader Gaussian component as component 2.
However, in the southwest region, the widths of the two components are similar, so we select the higher velocity, redshifted component as component 1 to maintain continuity in the velocity fields.
The narrower component is the systemic component, as its velocity field is broadly similar to that of the stellar velocity field from optical IFS stellar fitting \citep{davis12}.
The second component corresponds to the outflow as the position angle of the velocity field is similar to previous observations of the outflow \citep{alatalo11, davis12}, and the greater velocities and velocity widths are consistent with an outflow origin.
The outflow reaches velocities of 400 km/s, greater than the $\lesssim$100 km/s of the systemic component.
The width of the outflow component reaches linewidths up to 300 km/s, whereas the narrow systemic component remains around 75 km/s. 


\subsection{Br$\gamma$ Emission} \label{sec:brg}

In the central 100\,pc aperture, we measure a Br$\gamma$ flux of $6.8\times10^{-16}$erg/s/cm$^2$.
We apply an extinction correction from spatially unresolved Pa$\beta$ and Br$\gamma$ measurements from \citet{riffel13}.
We measure the visual extinction by assuming an intrinsic Pa$\beta$/Br$\gamma$ ratio of 5.89, corresponding to case B recombination, $n_e = 10^4$ cm$^{-3}$, and $T=10^4$ K \citep{osterbrock06}:
\begin{equation}
    A_V = 11.52 \log \left(\frac{5.89}{F_{Br\gamma, \text{obs}}/F_{Pa\beta, \text{obs}}} \right),
\end{equation}
yielding $A_V$ = 12.
We use the \citet{cardelli89} extinction curve with $R_V$ = 3.1, to find a Br$\gamma$ extinction of 1.4.
Thus, we obtain an extinction-corrected Br$\gamma$ flux of $2.4\times10^{-15}$erg/s/cm$^2$, or an extinction-corrected luminosity of $2.8\times10^{38}$erg/s.

We stress that this extinction correction is a lower limit.
\textit{Spitzer} observations show a silicate optical depth of $\tau_{9.7 \mu\text{m}} = 2.05$ \citep{smith07b, pellegrini13}, which corresponds to a V-band extinction of $A_V \sim$ 34 \citep{rieke85}.
This extinction measurement likely overestimates the extinction of the Br$\gamma$ emission, as some regions of the nucleus may be optically thick in the NIR and thus inflate the mean silicate optical depth, while the majority of the Br$\gamma$ emission could be from low-extinction regions.
Further, this silicate optical depth measurement is extracted over a roughly circular aperture with a radius of 10\arcsec, far larger than the field of view of the NIFS observations and may therefore be contaminated by excess extinction outside of the nuclear region.


\begin{table*}[!h]
    \centering
    \begin{tabular}{c|c|c|c|c|c}
    Line & Rest Wavelength & Center Flux & West Flux & East Flux & FOV Flux \\
     & $\mu$m & 10$^{-16}$ erg/s/cm$^2$ & 10$^{-16}$ erg/s/cm$^2$ & 10$^{-16}$ erg/s/cm$^2$ & 10$^{-16}$ erg/s/cm$^2$ \\ \hline
    H$_2$(1-0)S(2) & 2.0332 & 38.1 $\pm$ 0.6 & 16.01 $\pm$ 0.27 & 15.0 $\pm$ 0.15 & 298 $\pm$ 5 \\
    He I & 2.0587 & $<$50.0 & $<$19.64 & $<$19.45 & $<$346.0 \\
    H$_2$(2-1)S(3) & 2.0735 & 13.1 $\pm$ 0.3 & 3.99 $\pm$ 0.1 & 3.62 $\pm$ 0.14 & 74 $\pm$ 2 \\
    H$_2$(1-0)S(1) & 2.1213 & 96.8 $\pm$ 0.6 & 40.2 $\pm$ 0.2 & 40.52 $\pm$ 0.16 & 742 $\pm$ 5 \\
    H$_2$(2-1)S(2) & 2.1542 & 3.3 $\pm$ 0.2 & 0.82 $\pm$ 0.08 & 1.1 $\pm$ 0.12 & 16 $\pm$ 2 \\
    Br$\gamma$ & 2.1654 & 20.2 $\pm$ 0.8 & 1.24 $\pm$ 0.12 & 1.05 $\pm$ 0.1 & 44 $\pm$ 2 \\
    H$_2$(1-0)S(0) & 2.223 & 24.9 $\pm$ 0.7 & 8.96 $\pm$ 0.25 & 8.48 $\pm$ 0.17 & 177 $\pm$ 5 \\
    H$_2$(2-1)S(1) & 2.247 & 12.6 $\pm$ 0.7 & 4.54 $\pm$ 0.22 & 6.08 $\pm$ 0.17 & 107 $\pm$ 6 \\
    H$_2$(1-0)Q(1) & 2.4066 & 84.6 $\pm$ 0.7 & 13.08 $\pm$ 0.09 & $>$25.5 & $>$373 \\
    \end{tabular}
    \caption{Measured line fluxes for the three apertures shown in Figure~\ref{fig:flux_maps}. Fluxes reported here are the sum of the flux of both fitted components, when two components were used. The final column shows the measured fluxes over the entire field of view of the NIFS observations. Reported uncertainties do \textit{not} include the 10\% flux calibration uncertainty. Lower limits are indicated where part of the emission line falls out of the spectrum.}
    \label{tab:flux_table}
\end{table*}

\subsection{Ro-vibrational H$_2$ excitation diagrams}
\begin{figure}
    \centering
    \includegraphics[width=0.45\textwidth]{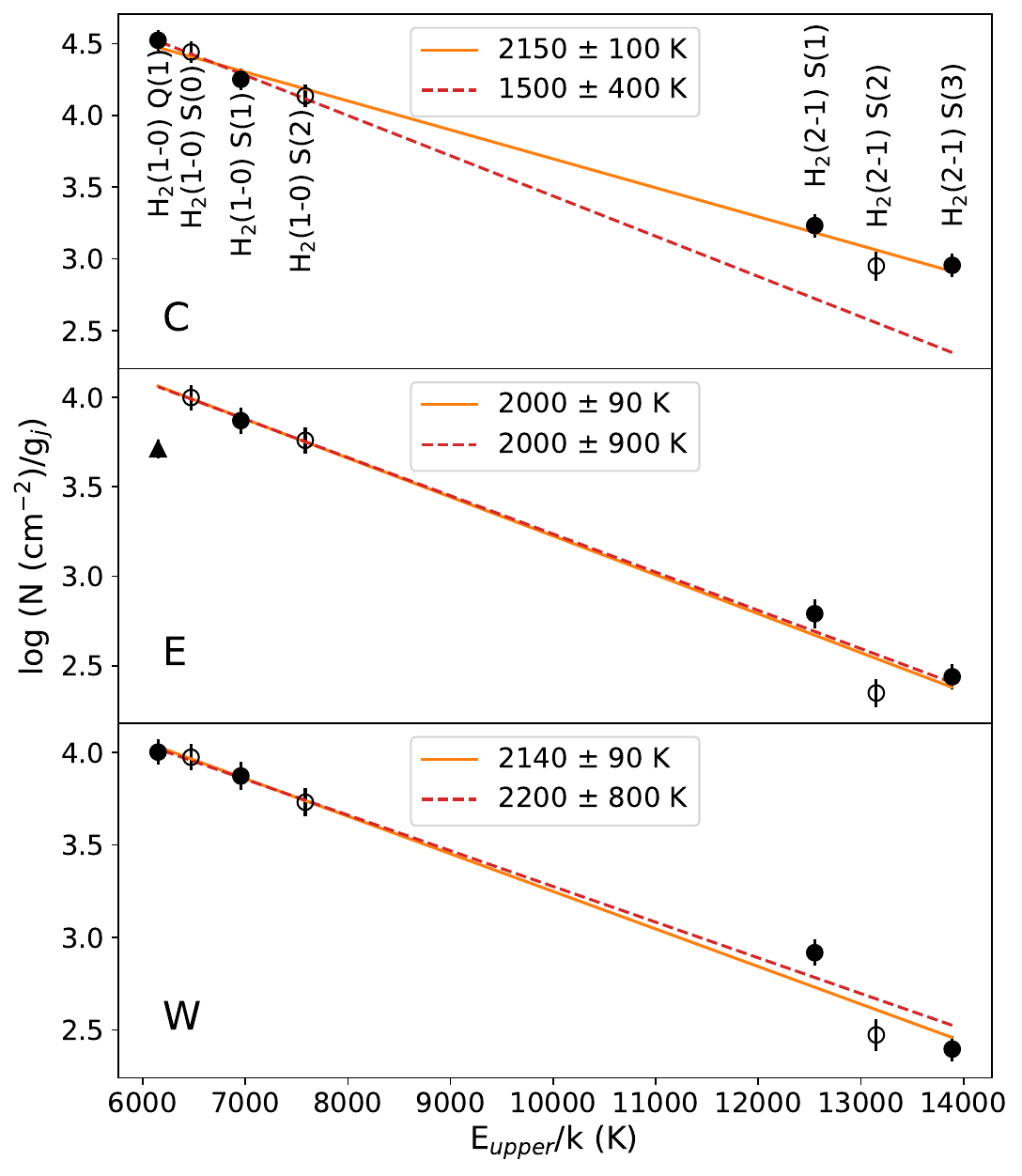}
    \caption{H$_2$ excitation diagram in the central, east, and west regions of NGC\,1266.
    Open and filled circles correspond to para and ortho lines, respectively. The upward triangle in the east region indicates a lower limit.
    The orange line shows a linear fit for all plotted emission lines, and the dashed red line is a linear fit to the (1-0) transition lines only. In the central aperture (top), the difference in slope between the red and orange fits shows that the (2-1) levels are relatively over-populated. 
    In the East and West apertures (middle and lower), a higher temperature of $\sim$2100 K is measured, and a single linear fit adequately describes the plotted transitions, indicating the gas is in LTE.}
    \label{fig:excitation}
\end{figure}

We can measure the excitation temperature of the hot H$_2$ traced by ro-vibrational emission. 
In Figure~\ref{fig:excitation} we plot the upper level excitation energy for each detected H$_2$ line against the column density normalized by the degeneracy for the transition.
We compute the column density following \citet{rosenberg13}:
\begin{equation}
    \frac{N_{\text{obs}}(\nu, J)}{g_J} = \frac{4\pi\lambda}{hc}\frac{I_{\text{obs}}(\nu, \nu', J, J')}{A(\nu, \nu', J, J')} \frac{1}{g_J},
\end{equation}
where $N_{\text{obs}}(\nu, J)$ is the upper energy level $(\nu, J)$ column density, $g_J$ is the degeneracy of this level, $I_{\text{obs}}(\nu, \nu', J, J')$ is the observed flux of the transition, and $A(\nu, \nu', J, J')$ is the associated Einstein radiative transition probability from \citet{wolniewicz98}.
For thermally excited gas in local thermodynamic equilibrium (LTE), the column densities in the excitation diagram will be linearly related to the energies with a single temperature \citep[e.g.][]{davies03}.

To measure the excitation temperature of the molecular hydrogen, we linearly fit the excitation diagram considering all detected emission lines, and the (1-0) lines separately.
For the central aperture, the combined linear fit yields a temperature of 2150 $\pm$ 100 K, while the fit to the (1-0) lines only has a steeper slope with a temperature of 1500 $\pm$ 400 K.
Despite the large uncertainty on the (1-0) only fit, the difference in slope indicates that the (2-1) energy levels are overexcited compared to LTE conditions.
Thus, the nucleus likely contains gas at multiple temperatures that cannot be adequately described with a single temperature, where the lower energy (1-0) levels are weighted towards lower temperatures.
In the West and East apertures, the combined linear fit and the fit to only the (1-0) lines are similar, with excitation temperatures ranging from $2000-2200$ K. 

\subsection{Ortho-to-Para Ratio}
The ortho-to-para ratio (OPR) can be used as a diagnostic to aid in determining the excitation source of warm H$_2$.
This ratio measures the amount of emission from ortho-H$_2$ molecules (i.e. where the nuclear spins are aligned) to that of para-H$_2$ (i.e. the spins are anti-aligned).

We measure the ro-vibrational OPR in each aperture following \citet{harrison98}.
In the central, west, and east apertures we find ortho-to-para ratios of 2.6 $\pm$ 0.4, 2.8 $\pm$ 0.4, and  3.0 $\pm$ 0.5  respectively for the $\nu$=1 transitions (i.e. (1-0) S(1)/(1-0) S(0)).
All of these values are consistent with the equilibrium value of 3 within uncertainties.
While an equilibrium value of the OPR typically indicates that the excitation source is thermal, it is possible for a dense PDR to have an OPR of 3 due to collisional de-excitation \citep{sternberg89, burton90}.

\subsection{H$_2$ Line Diagnostics}
\begin{figure}
    \includegraphics[width=0.47\textwidth]{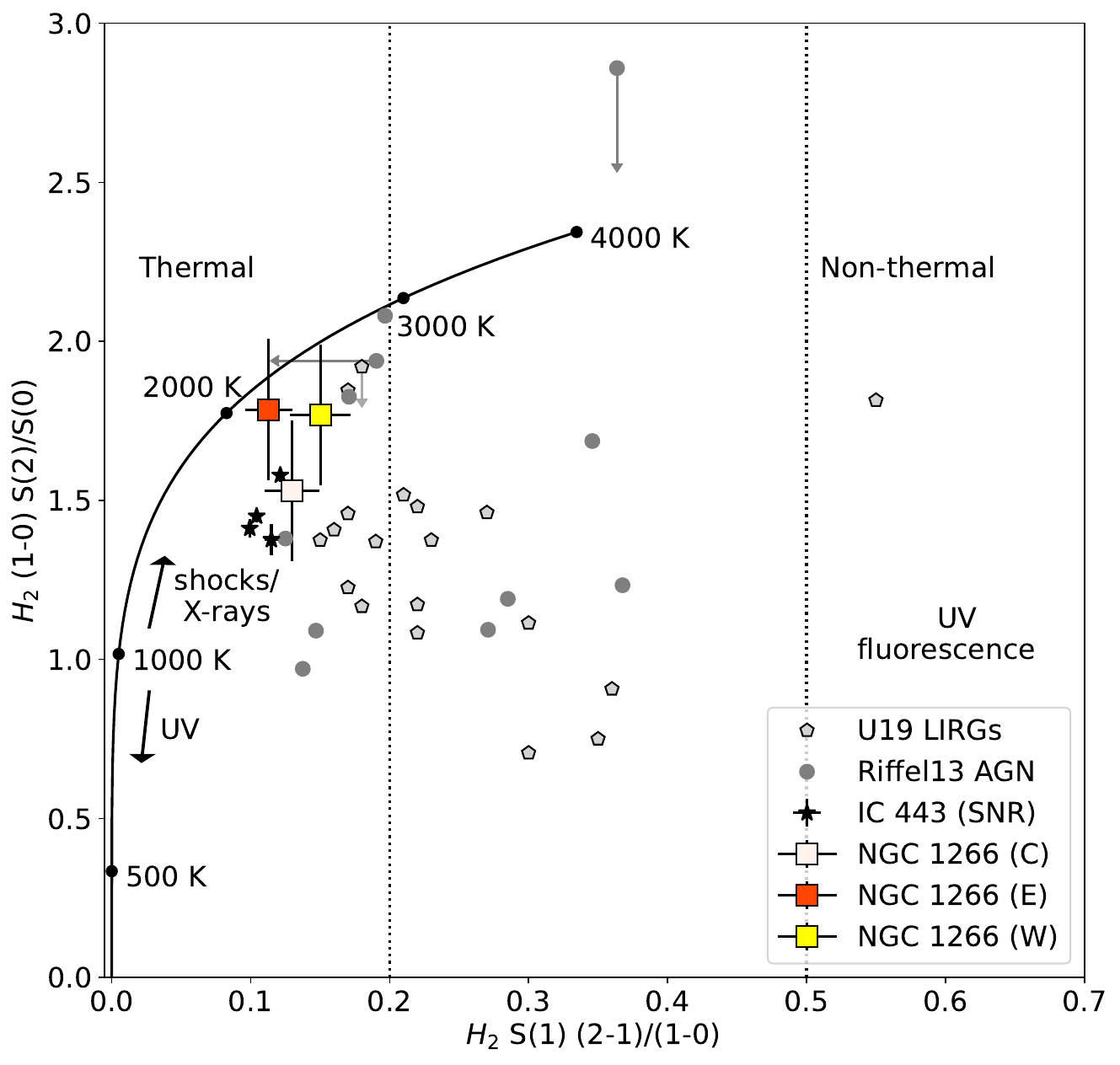}
    \caption{H$_2$ line ratio plot for extracted apertures of NGC\,1266 and comparison samples. The orange, white, and yellow squares show the ratios in the three extracted apertures shown in Figure~\ref{fig:flux_maps}. The two outer apertures have similar H$_2$ line ratios, while the central aperture is farther from the black line showing the expected LTE line ratios at the labelled temperatures. UV photons account for excitation below 1000 K while shocks and X-rays heat the gas above 1000 K. The non-thermal region on the right side shows line ratios for UV fluorescence from \citet{black87}. The black stars show the line ratios for IC 443 from \citet{deng23}, a shocked supernova remnant. The outlined pentagons show measurements from a sample of LIRGs  from \citet{u19a}, and the gray circles are a sample of AGN from \citet{riffel13}.}
    \label{fig:h2_ratio}
\end{figure}

Various ratios of H$_2$ ro-vibrational lines provide useful diagnostics for determining the excitation mechanism of the gas.
Figure~\ref{fig:h2_ratio} plots two line ratios: (1-0)S(2)/(1-0)S(0) on the y-axis (a ratio of para transitions); and (2-1)S(1)/(1-0)S(1) on the x-axis (a ratio of ortho transitions).
The ortho ratio is useful for differentiating between thermal, non-thermal, and mixed excitation mechanisms, whereas the para ratio is driven upward with increasing excitation temperature.
We note that in a dense PDR, collisional de-excitation can drive the (2-1)S(1)/(1-0)S(1) ratio lower, and thus, it appears to be thermally excited \citep{harrison98}.
We plot our three apertures in colored points, and additionally plot these line ratios for a sample of AGN \citep[dark gray circles][]{riffel13} and Luminous InfraRed Galaxies \citep[LIRGs, light gray pentagons][]{u19a} for reference.
We also include measurements for a shocked supernova remnant, IC\,443 \citep{deng23}.
The black line shows the expected line ratios for a gas of uniform density in thermodynamic equilibrium.
The two outer apertures of NGC\,1266 are consistent with the purely thermally excited molecular gas at approximately 2000 K, as we might expect from shock excitation.
The central aperture has a lower H$_2$(1-0) S(2)/S(0) ratio, and is thus closer to bulk of the AGN and LIRG samples, indicating a possible photo-dissociation region (PDR) contribution to the excitation in the nucleus from either an AGN or starburst as would be expected in AGN hosts and LIRGs.

\begin{figure}
    \includegraphics[width=0.47\textwidth]{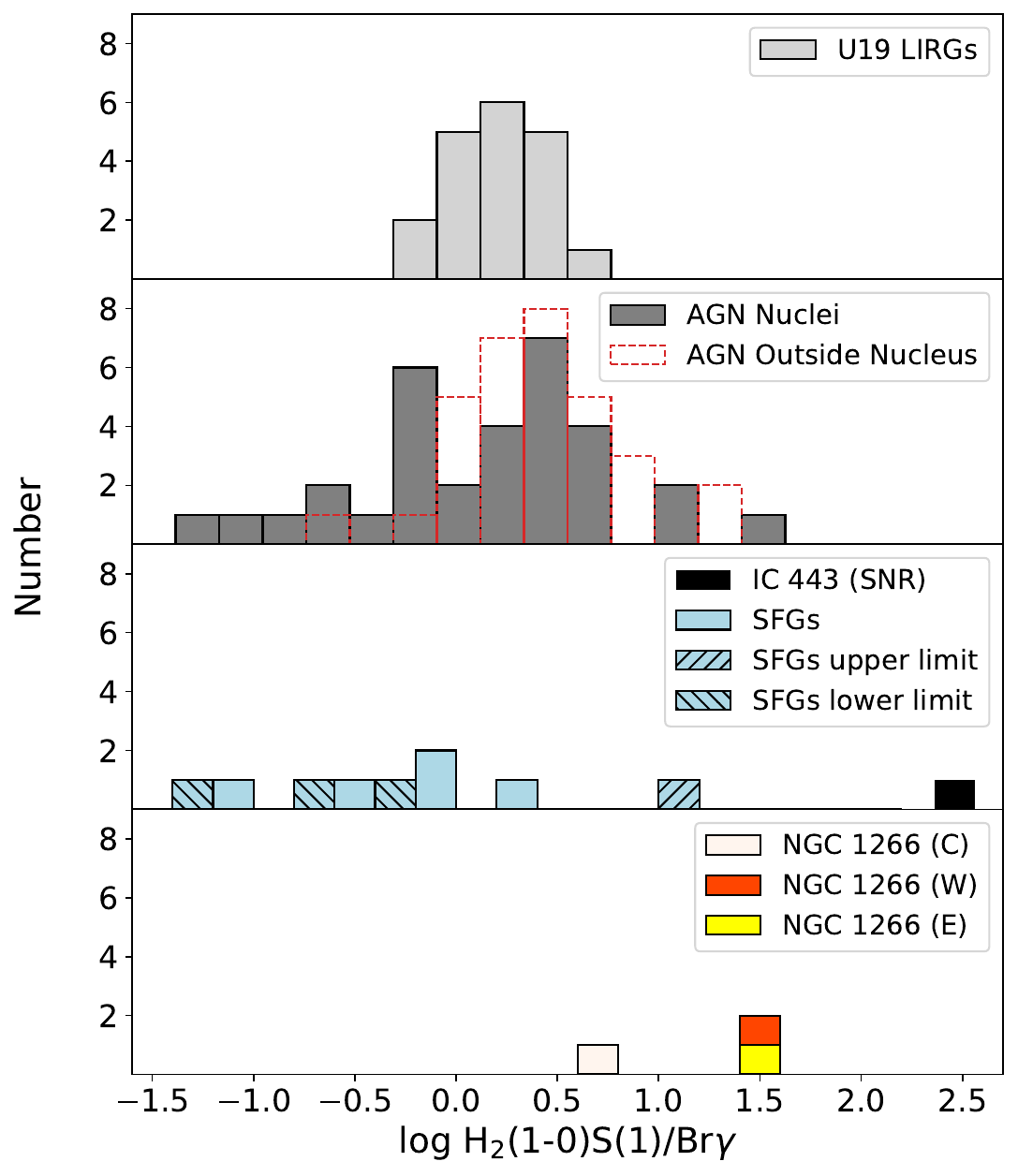}
    \caption{Histograms of the log H$_2$(1-0)S(1)/Br$\gamma$ ratio for various samples. The top panel shows the LIRG sample from \citet{u19a}. The next panel is the AGN sample from \citet{riffel21}. The AGN nuclei are plotted in dark grey bars, where the nuclei measurements are within the central PSF FWHM (mean of 60\,pc). The red dashed outline shows the measurements from the same sample but outside the nuclear region plotted in gray. The third panel shows a sample of star-forming galaxies from \citet{dale04} in light blue bars. Diagonally hatched bars show upper and lower limit measurements. We also show the measurement from IC\,443, a shocked supernova remnant \citep{deng23}. In the bottom panel, we show the measurements from the three NGC\,1266 apertures. The East aperture in yellow is a lower limit as Br$\gamma$ is not detected.}
    \label{fig:h2_brg}
\end{figure}

Another important NIR diagnostic is shown in Figure~\ref{fig:h2_brg}, the H$_2$(1-0)S(1)/Br$\gamma$ ratio (hereafter H$_2$/Br$\gamma$ ratio).
This ratio is useful for constraining the excitation mechanism by tracing the relative amounts of ionized gas and warm molecular gas. 
Star-forming galaxies tend to exhibit H$_2$/Br$\gamma$ $<$ 0.4, AGN lie in the range 0.4 $<$ H$_2$/Br$\gamma$ $<$ 6, and shocked regions have ratios $>6$ \citep{riffel13}. 
In the center of NGC\,1266, we find an H$_2$/Br$\gamma$ ratio of 4.8 $\pm$ 0.7, whereas the East and West apertures have ratios of 32 $\pm$ 6 and 39 $\pm$ 7, respectively.
While the H$_2$ line ratios discussed above have little spatial variation, the H$_2$/Br$\gamma$ varies significantly due to the centrally concentrated nature of the Br$\gamma$ emission, as shown in Figure~\ref{fig:flux_maps}.
Comparing to other observational samples of galaxies, the center of NGC\,1266 is consistent with the plotted LIRG and AGN samples, while the East and West measurements have values beyond nearly all other observed galaxies, and consistent with shock excitation.

\subsection{Warm molecular gas masses}
We compute the warm molecular gas mass from our H$_2$(1-0)S(1) flux following \citet{mazzalay13}.
We assume a temperature $T=2000$ K (as is consistent with our measurements in Figure~\ref{fig:excitation}), an H$_2$(1-0)S(1) transition probability $A_{S(1)} = 3.47\times10^{-7}$ s$^{-1}$, and a population fraction in the $(\nu, J) = (1, 3)$ level of 0.0122.
This yields the relation:
\begin{equation} \label{eq:mass}
    M_{H_2, \text{warm}} \simeq 5.0875 \times 10^{13} \left(\frac{D}{\text{Mpc}}\right)^2 \left(\frac{F_{(1-0)S(1)}}{\text{erg s}^{-1} \text{cm}^{-2}} \right) 10^{0.4 A_{2.2}}
\end{equation}

\noindent where $D = 29.9$ Mpc is the distance to NGC\,1266, $F_{(1-0)S(1)}$ is the flux of the H$_2$(1-0)S(1) line, and $A_{2.2}$ is the extinction at 2.2 $\mu$m.
To derive $A_{2.2}$, we use the measured $A_V$ from \citet{riffel13}, as in Section~\ref{sec:brg}, and adopt the same $R_V = 3.1$.
We report our measured masses in Table~\ref{tab:meas_table}.
We again note that the extinction derived in Section~\ref{sec:brg} from the Pa$\alpha$ to Br$\gamma$ is likely an underestimate of the true extinction; thus these measured gas masses should be considered lower limits.



\begin{table*}
    \centering
    \begin{tabular}{c|c|c|c|c}
      & Center & West & East & Entire FOV \\ \hline
    Warm H$_2$ mass ($M_\odot$)$^{a}$ & 1540 $\pm$ 150 & 640 $\pm$ 60 & 650 $\pm$ 60 & 11,800 $\pm$ 1,200 \\
    Ortho-to-para ratio$^{b}$ & 2.6 $\pm$ 0.4 & 2.8 $\pm$ 0.4 & 3.0 $\pm$ 0.5 & 2.7 $\pm$ 0.4\\
    H$_2$/Br$\gamma$ ratio$^c$ & 4.8 $\pm$ 0.7 & 32 $\pm$ 6 & 39 $\pm$ 7 & 17 $\pm$ 3 \\
    $L_{Br\gamma}$ ($10^{38}$ erg/s)$^{d}$ & 8.4 $\pm$ 0.9 & 0.52 $\pm$ 0.07 &  0.44 $\pm$ 0.06 &  18 $\pm$ 2 \\
    SFR (M$_\odot$/yr)$^{e}$ & 0.69 $\pm$ 0.07 & 0.042 $\pm$ 0.005 & 0.036 $\pm$ 0.005 & 1.51 $\pm$ 0.17 \\
    \end{tabular}
    \caption{Computed quantities in each aperture and across the entire field of view of the NIFS observations. Reported uncertainties include the 10\% flux calibration uncertainty. (a) measured warm H$_2$ masses calculated according to Equation~\ref{eq:mass}. (b) H$_2$ $\nu$=1 ortho-to-para ratio. (c) ratio of H$_2$(1-0)S(1)/Br$\gamma$. (d) dust-corrected Br$\gamma$ luminosity. (e) dust-corrected Br$\gamma$ star-formation rate, assuming all Br$\gamma$ is from star-formation.}
    \label{tab:meas_table}
\end{table*}

\section{Discussion} \label{sec:discuss}

\subsection{H$_2$ Excitation Mechanisms}

Overall, our results demonstrate that the East and West regions are thermally excited by shocks while we find evidence of an additional excitation mechanism in the nucleus.
The excitation diagrams of the East and West apertures (see Figure~\ref{fig:excitation}) are well-fit by a single temperature component of $\sim$2100 K, consistent with thermal excitation at a single temperature.
In the nucleus, the increase in temperature when fitting all detected H$_2$ transitions instead of only the (1-0) transitions indicates that the gas is not well-described by a single temperature component because the (2-1) transitions are overpopulated relative to the (1-0) transitions.
The lack of a single temperature component may indicate that there is either a contribution to the emission line flux from a non-thermal excitation mechanism such as UV fluorescence, or multiple temperature components are present.

From comparisons with other galaxies, the over-populated (2-1) transitions in the nucleus would be consistent with an excitation contribution from either AGN emission or star-formation.
Similar ro-vibrational excitation diagrams have been observed in both starburst and AGN-host galaxies \citep{rosenberg13, emonts14}. 
The nuclear excitation diagram of NGC\,1266 is especially similar to the excitation diagram of the southern, obscured nucleus of NGC\,3256 -- a dual-AGN host galaxy with a likely AGN-driven outflow \citep{emonts14}.
In that object, both nuclei have similar excitation temperatures of $\sim$2000 K when fitting all transitions, and $\sim$1600 K when only fitting the (1-0) transitions.
\citet{emonts14} conclude that the molecular gas in NGC\,3256 is likely heated to the observed temperature of 2000 K by either slow shocks or X-rays (potentially originating from the AGN).
In NGC\,1266, a contribution from both shocks and X-rays from the AGN could explain the observed departure from a single excitation temperature observed in shock-dominated systems.

Line ratio diagnostics further support a dominant thermal excitation mechanism heating the gas to a single excitation temperature in the outskirts, with a non-shock contribution in the nucleus.
In the H$_2$ ratio diagram shown in Figure~\ref{fig:h2_ratio}, purely thermal excitation has (2-1)S(1)/(1-0)S(1) $\lesssim$ 0.2, and non-thermal has (2-1)S(1)/(1-0)S(1) $\gtrsim$ 0.5 \citep{black87, brand89, draine90, mouri94, dors12}.
The ratio of para transitions, (1-0)S(2)/(1-0)S(0) increases with higher excitation temperatures and can distinguish between thermal UV excitation and shock or X-ray excitation as UV excitation does not heat the molecular gas above $\sim$1000 K \citep{sternberg89}.
The West and East NGC\,1266 apertures lie along the expected LTE relation above 2000 K, while the central aperture is offset below this line and is closer to the majority of the LIRG sample of \citet{u19a}. 
The outer apertures thus are consistent with shock or X-ray excitation while offset below the LTE relation is consistent with a non-shock contribution in the center. 
In comparison to the AGN sample of \citet{riffel13} and the LIRG nuclei sample of \citet{u19a}, NGC\,1266 occupies a relatively unpopulated region of this diagram, where the many galaxies in these other samples appear to have some non-thermal contribution to their H$_2$ excitation, likely from UV fluorescence originating from AGN or star-formation.
The high H$_2$/Br$\gamma$ ratio shown in Figure~\ref{fig:h2_brg} further supports shock excitation in the East and West apertures outside the nucleus, while the center has some star-formation or AGN contribution.

Finally, the observed H$_2$ ortho-to-para ratio of 3 is consistent with a thermal excitation mechanism including shocks.
Hydrogen molecules can be excited into different quantum states through reactive collisions with protons or Hydrogen atoms.
While the OPR is initially set by H$_2$ formation mechanisms, once the gas reaches LTE and has $T > 200$K, the equilibrium OPR is 3.
However, if the H$_2$ is excited from FUV-pumping and only vibrationally excited states (i.e. $\nu \geq 1$) are considered, the equilibrium OPR is 1.7 \citep{sternberg89}.
In a PDR where FUV pumping is present, the observed OPR is expected to lie near this value. 
However, in a dense PDR, collisional de-excitation of H$_2$ can drive the OPR to the equilibrium value.
K-band observations of starburst galaxies have yielded sub-equilibrium OPR values, indicating that the excitation has contributions both from a thermal source (shocks or UV irradiation) and FUV pumping \citep{harrison98, davies03}.
Our observed OPRs of 3 therefore do not disqualify the possibility of excitation from a dense PDR, but are consistent with our conclusion that the H$_2$ is primarily excited by shocks.

\subsection{Shock Diagnostics}

Insight into the properties of the shock in NGC\,1266 is crucial to understanding how the shock is propagating through the galaxy and the impact it has on the ISM.
Previous shock modelling of expected H$_2$ ro-vibrational and Br$\gamma$ emission can help characterize the shock properties in NGC\,1266.
Shocks are typically described as either continuous (C-shocks) or jump (J-shocks), where continuous shocks are modelled with continuously varying properties such as temperature and density and jump shocks have a discontinuous change in properties at the shock front.
NIR spectral properties of four classes of shocks can be summarized as the following \citep{burton90, davies00}.
\begin{enumerate}[label=(\roman*)]
    \item \textit{Fast J-shocks} ($v$ from 100-300 km/s) are highly dissociative, resulting in similar Br$\gamma$ and H$_2$(1-0)S(1) fluxes.
    \item \textit{Slow J-shocks} ($v$ from 5-15 km/s) are not dissociative resulting in strong H$_2$ emission and weak Br$\gamma$, and high post-shock gas temperatures around 4000 K.
    Typically these shocks will revert to a C-shock \citep{hollenbach89}.
    \item \textit{Fast C-shocks} ($v$ from 25 - 40 km/s) would heat the gas from 1000-3000 K depending on the shock velocity, resulting in strong H$_2$ emission and little Br$\gamma$ as these shocks are not strongly dissociative.
    \item \textit{Slow C-shocks} ($v \leq 15$ km/s) only heat the gas to $\sim$300 K with weak H$_2$ emission.
\end{enumerate}
Thus, our observed gas temperature of $\sim$2000 K and high $H_2$/Br$\gamma$ ratio is most consistent with a fast C-shock.

Previous studies of the shocks in NGC\,1266 have relied on MIR H$_2$ pure rotation lines and the CO and H$_2$O spectral line energy distributions, and find that a C-shock with n$_H$=10$^4$\,cm$^{-3}$ and $v_s$=$25-30$\,km/s is most consistent with these observations \citep{pellegrini13, chen23}.
Both analyses require a contribution from PDRs in addition to shocks to reproduce the observed fluxes, which is expected as the observations were not spatially resolved and thus had no differentiation between the central core and the surrounding region.
Comparing to NIR shock models, we find that these derived shock properties from the MIR are consistent with our observed NIR H$_2$ and Br$\gamma$ emission.

\subsection{Resolved shock structure}

In Figure~\ref{fig:flux_maps}, we resolve the structure of the shock in the central 450\,pc of NGC\,1266 with an unambiguous shock tracer for the first time.
The H$_2$ flux shows a clear structure extending from the northwest to the southeast.
While the morphology of the H$_2$ emission may have implications for the physical mechanism driving the shocks (i.e. radio jets or supernovae), it is difficult to make conclusions without accompanying multi-wavelength observations at similar spatial resolution.
Forthcoming high-resolution VLA observations of NGC\,1266 will be presented in Otter et. al. (in prep.) and may clarify whether the shocked molecular gas is a product of jet-ISM interaction.

\subsection{Star-formation or AGN?}

Our results show that the excitation source in the central 100\,pc is different from the surrounding region.
While the outskirts show clear signs of shock excitation, the center appears to have some contribution from another mechanism as the (2-1) levels are elevated above the thermalized temperature of the (1-0) levels (see Figure~\ref{fig:excitation}).
Moreover, the central region has an H$_2$/Br$\gamma$ ratio consistent with AGN nuclei and LIRGs in Figure~\ref{fig:h2_brg}, while Br$\gamma$ is either undetected or weak in the East and West spectra.
Either AGN emission or central star-formation may be responsible for the Br$\gamma$ emission.

To unambiguously determine whether the central energy source is dominated by the AGN or star-formation, a reliable measurement of the AGN bolometric luminosity is needed.
\citet{alatalo15} first estimate the AGN bolometric luminosity by extrapolating from the IR luminosity of the compact central source and a full SED fit, finding $L_{bol}=3.4\times10^{43}$\,erg/s.
However, a reanalysis of the full SED by \citet{chen23} finds a lower AGN contribution to the IR luminosity, and a resulting lower total luminosity of $L_{bol}=1.25\times10^{42}$\,erg/s.

We can roughly estimate the AGN contribution to the central Br$\gamma$ emission for these differing $L_{bol}$ measurements using the empirical $L_{bol}$-$L_{H\alpha}$ relation \citep{greene07a}, and then converting the expected H$\alpha$ luminosity to Br$\gamma$ by assuming Case B recombination and the same gas properties as in Section~\ref{sec:brg}.
Using the \citet{alatalo15} $L_{bol}$, the expected Br$\gamma$ luminosity from the AGN is 10$^{39}$\,erg/s, which is slightly higher than our measured Br$\gamma$ luminosity.
If the \citet{alatalo15} measure of $L_{bol}$ is correct, this discrepancy could be due to an under-estimate of the dust extinction from the lower spatial resolution \citet{riffel13} observations.

Alternatively, if we adopt the $L_{bol}$ from \citet{chen23}, we find that the AGN contributes little to the Br$\gamma$ emission, and thus that the Br$\gamma$ emission is almost entirely due to central star-formation.
For $L_{bol}=1.25\times10^{42}$\,erg/s, the expected Br$\gamma$ luminosity is 2.2$\times$10$^{37}$\,erg/s, well over an order of magnitude \textit{below} our measured Br$\gamma$ luminosity.
In this scenario, we assume the Br$\gamma$ emission is purely star-formation and measure a nuclear star-formation rate of 0.69 $\pm$ 0.07\,M$_\odot$/yr using the adjusted \citet{kennicutt98} H$\alpha$ star-formation law presented by \citet{u19a}, though we note that we may be underestimating the dust extinction, as discussed in Section~\ref{sec:brg}.

While a variety of multi-wavelength star-formation tracers yield measurements spanning an order of magnitude, this Br$\gamma$ measurement constrains the on-going star-formation on the timescale of the outflow. 
The outflow appears to have originated $\lesssim$5 Myr ago \citep{davis12}, and with a recent burst of star-formation within the last $\sim$500\,Myr, tracing the star-formation on short timescales is crucial for identifying the impact of the outflow on the star-formation.
\citet{alatalo15} perform an exhaustive study of star-formation tracers in NGC\,1266, finding a range of values from $\sim$0.3 M$_\odot$/yr to $\sim$2.2 M$_\odot$/yr.
While such a discrepancy may appear alarming, every individual star-formation tracer has a unique combination of contaminants (primarily AGN emission and shocks), suppressants (e.g. dust extinction, dust destruction), and crucially, trace star-formation on different timescales.

A variety of star-formation tracers in other post-starburst galaxies also yield a wide range of values likely due to differing timescales probed and contamination \citep{smercina18, luo22}. 
Star-formation tracers that originate from ionized HII regions surrounding young stars trace star-formation on timescales of order the lifetime of massive stars, $\sim$10 Myr \citep[e.g.][]{ostriker74}.
Hydrogen recombination lines and forbidden emission lines in particular provide an excellent probe of existing HII regions, though they can be significantly contaminated by shocks.
\citet{alatalo15} measure the star-formation rate of NGC\,1266 with the mid-infrared [NeII] 12 $\mu$m and [NeIII] 15 $\mu$m emission lines of $\approx$1.5 M$_\odot$/yr.
While MIR emission lines have the advantage of little dust extinction, this derived star-formation rate may be heavily contaminated by shocks or the AGN, and the \textit{Spitzer} observations these measurements are made from lack the spatial resolution to resolve the morphology of this emission.

The most robust star-formation tracer from \citet{alatalo15} is the free-free emission originating from HII regions, yielding a star-formation rate of 0.87 M$_\odot$/yr, consistent with our measurement within uncertainties assuming all Br$\gamma$ emission is from star-formation.
The free-free measurement has the advantage of having little dust extinction, though it lacks spatial information as the full SED fitting required to derive the free-free contribution uses photometry from observations with a variety of angular resolutions.
We stress that each star-formation tracer provides different physical information and is subject to significant uncertainties, and that multiple tracers should be considered when constraining the star-formation rate.
High-resolution mid-infrared and radio observations are needed to progress in disentangling the multiple energetic processes, unusual stellar population, and high extinction in this galaxy.

\begin{figure}
    \centering
    \includegraphics[width=0.47\textwidth]{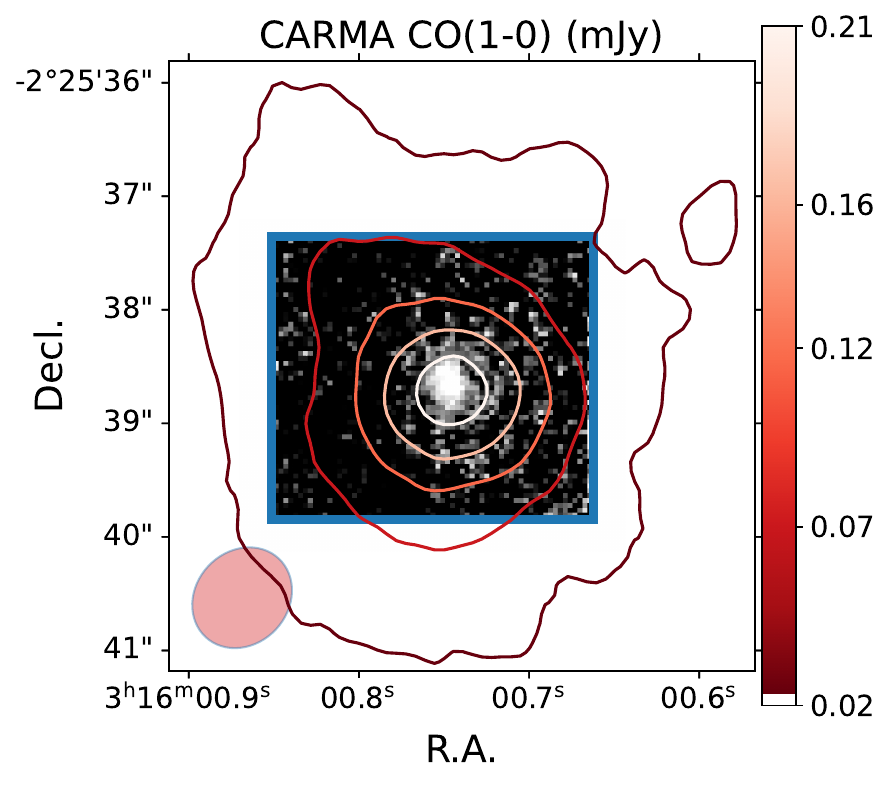}
    \caption{Br$\gamma$ flux map with the NIFS field of view shown with the blue rectangle. The red contours show CARMA CO(1-0) observations from \citet{alatalo15}. The red ellipse shows the CARMA beam size. Though at coarser resolution than the NIFS data, the CO extent is much larger than the Br$\gamma$ emission.}
    \label{fig:co_map}
\end{figure}

We compare the Br$\gamma$ emission morphology to that of the CARMA CO(1-0) observations from \citet{alatalo15} in Figure~\ref{fig:co_map}.
We see that the CO emission extends far beyond the Br$\gamma$ emission, consistent with scenario 2 in \citet{alatalo15} where the majority of the molecular gas lacks significant star-formation and the existing star-formation is concentrated in a nuclear starburst.
While it is unclear whether the source of the Br$\gamma$ emission is this nuclear starburst or AGN, there is little star-formation in the the surrounding CO-emitting region of NGC\,1266, indicating the vast majority of the molecular gas hosts suppressed star-formation.

\subsection{Implications for star-formation quenching}
Our results show that NGC\,1266 hosts a large reservoir of warm, shocked molecular gas in the central region with an extended shock morphology.
We also find that the central 400\,pc region has very little star-formation, with the possible exception of a $\lesssim$100\,pc nuclear star cluster.

\citet{alatalo15} propose two scenarios for star-formation in NGC\,1266, a nuclear starburst or distributed low-level star-formation throughout the disk.
Our results are thus consistent with the former, though it remains unclear precisely \textit{why} this star-formation is constrained only to the very nucleus and star-formation suppressed elsewhere.
This suppression is not solely due to a lack of cold molecular gas, as NGC\,1266 hosts a significant reservoir of CO(1-0) over a $\sim$1\,kpc extent \citep{alatalo11}.
Though we find a significant amount of shock heated molecular gas, \citep{alatalo15} detect multiple dense gas tracers such as CS(2-1) over a $\sim$500\,pc extent, indicating that star-formation is not suppressed due to a lack of available dense gas.

\citet{alatalo15} suggest that the outflow drives turbulence throughout the ISM of NGC\,1266, effectively suppressing star-formation.
In this scenario, our results indicate that this star-formation suppression is efficient, as the star-formation in NGC\,1266 is mostly relegated to the nucleus.
In simulations and observations, excess turbulence has been shown to suppress star-formation by increasing the virial parameter (i.e. the ratio between twice the kinetic to gravitational energy) in molecular clouds, thus making gravitational collapse more difficult.
While turbulence can locally enhance star-formation through local compression of gas, over larger volumes the increase in virial parameter has a greater impact, thus suppressing the star-formation overall \citep{federrath12, mandal21a, federrath16, salim20}.

Turbulent stabilization of the molecular gas may be a necessary component of star-formation suppression mechanisms that do not rely on exhausting or expelling the molecular gas reservoirs.
A number of studies have revealed that many post-starburst galaxies harbor significant cold molecular gas reservoirs despite their lack of star-formation \citep{french15, rowlands15, otter22}.
High turbulence levels have been observed in a few post-starbursts, though the origin of this turbulence is unclear \citep{smercina22}.
Outflows are often observed in PSBs, further supporting that AGN feedback through outflows may play an important role in the evolution through the post-starburst phase and beyond \citep[e.g.][]{otter22, smercina22, luo22, french23}.

\section{Conclusions} \label{sec:conclusion}
In this work, we present K-band IFS observations of the central 500\,pc of NGC\,1266 yielding ro-vibrational H$_2$ emission maps, the first unambiguous shock tracer showing the morphology of the shock in this galaxy and allowing us to study the NIR nucleus at spatial high resolution, probing physical scales of $\lesssim$50 pc.
We detect 7 ro-vibrational H$_2$ emission lines and Br$\gamma$.

We make the following conclusions:
\begin{enumerate}
    \item The H$_2$ ro-vibrational emission is primarily excited by a high velocity C-shock, heating the molecular gas to approximately 2000 K, and the H$_2$ excitation properties are relatively uniform across the central $\sim$400 pc.
    \item We have observed the morphology of the shock in the central region for the first time, though multi-wavelength high spatial resolution observations are needed to determine the source of the shock.
    \item The Br$\gamma$ and H$_2$ excitation in the nucleus contains some contribution from a photo-dissociation region originating from the AGN, star-formation or a combination of both.
    \item If the Br$\gamma$ emission in the nucleus is dominated by star-formation, we measure a central star-formation rate of 0.69 $\pm$ 0.07 M$_\odot$/yr (only including the formal uncertainty), consistent with previous short-timescale measurements of ongoing star-formation. We note that the dust extinction may be higher than our estimate, and thus this star-formation rate should be considered a lower limit, assuming little AGN contamination.
    \item In either case, we observe little non-nuclear star-formation in NGC\,1266, indicating that the vast majority of the molecular gas has its star-formation suppressed, potentially by turbulence driven by the outflow. 
\end{enumerate}

The multitude of physical processes ongoing in NGC\,1266 make it an ideal laboratory for studying AGN feedback, star-formation suppression, extragalactic shocks, jet-ISM interaction, and most importantly, how all of these processes are intertwined.
With the NIR observations presented here, we begin to resolve the structure of the shock for the first time, as is essential for uncovering the physics connecting the radio jets, to the outflows, to the shocks, and to the star-formation.
This work highlights the importance of spatially resolved studies of shock and star-formation properties, as the excitation and gas properties in the central 100\,pc are complex.
Moving forward, high spatial resolution NIR and MIR observations over a large field of view with \textit{JWST} are essential for fully characterizing the shock properties, structure, and the role of outflow-driven turbulence in star-formation suppression. 
Further in the future, new radio telescopes with advancements in sensitivity and resolution, such as the next-generation Very Large Array, will extend the work presented here to higher redshift sources and enable statistical studies of the role of jet-ISM interactions in galaxy evolution \citep[e.g.][]{nyland18a}.

\section{Acknowledgements}

We thank the anonymous referee for their comments which have improved this work.
We would like to thank Michelle Cluver for her contributions to obtaining the Gemini-NIFS observations.

J.~A.~O.~acknowledges support from the Space Telescope Science Institute Director's Discretionary Research Fund grants D0101.90296 and D0101.90311. Y.~L.~, A.~P.~ and P.~P~acknowledge support from SOFIA grant \#08-0226 (PI: Petric). T.A.D acknowledges support from the UK Science and Technology Facilities Council through grants ST/S00033X/1 and ST/W000830/1. Y.~L.~acknowledges support from the Space Telescope Science Institute Director's Discretionary Research Fund grant D0101.90281. 
P.~P~, K.~R. and K~.A.~acknowledge support from NASA grant 80NSSC23K0495. M.~S.~acknowledges support from the William H. Miller III Graduate Fellowship.
C.~F.~acknowledges funding provided by the Australian Research Council (Discovery Project DP230102280), and the Australia-Germany Joint Research Cooperation Scheme (UA-DAAD).
S.~A. ~~acknowledges support from the European Research Council (ERC) under the European Union’s Horizon 2020 research and innovation programme (grant agreement No 789410. S.~A. and G.~O. acknowledges funding from the Knut and Alice Wallenberg foundation (KAW). 

Based on observations obtained at the international Gemini Observatory, a program of NSF NOIRLab, acquired through the Gemini Observatory Archive at NSF NOIRLab* and processed using the Gemini IRAF package, which is managed by the Association of Universities for Research in Astronomy (AURA) under a cooperative agreement with the U.S. National Science Foundation on behalf of the Gemini Observatory partnership: the U.S. National Science Foundation (United States), National Research Council (Canada), Agencia Nacional de Investigaci\'{o}n y Desarrollo (Chile), Ministerio de Ciencia, Tecnolog\'{i}a e Innovaci\'{o}n (Argentina), Minist\'{e}rio da Ci\^{e}ncia, Tecnologia, Inova\c{c}\~{o}es e Comunica\c{c}\~{o}es (Brazil), and Korea Astronomy and Space Science Institute (Republic of Korea).
Basic research in radio astronomy at the U.S. Naval Research Laboratory is supported by 6.1 Base Funding. 

This work was enabled by archival observations made from the Gemini North telescope, located adjacent to the summit of Maunakea.
We acknowledge that Hawai'i and Maunakea are unceded territory of the Kanaka Maoli people and support their struggle for liberation and self determination from colonial powers. 

\facilities{Gemini:Gillett, CARMA, HST}
\software{astropy \citep{astropycollaboration13, astropycollaboration18}, scipy \citep{virtanen20}, numpy\citep{harris20}, matplotlib \citep{hunter07}}

\appendix

\section{H$_2$ Flux Maps}

\begin{figure*}
    \centering
    \includegraphics[width=0.8\textwidth]{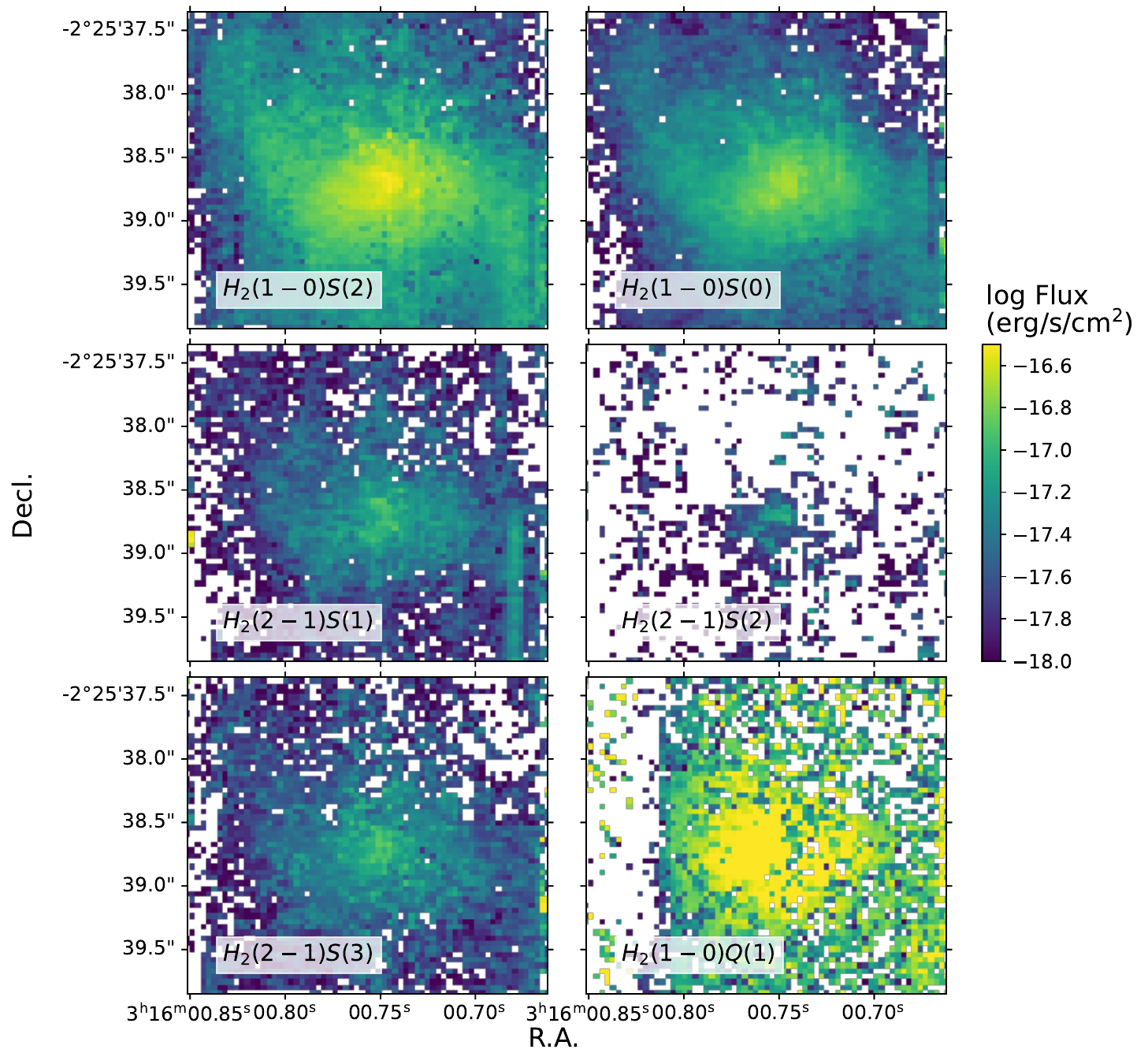}
    \caption{Flux maps for all detected H$_2$ emission lines. Line fluxes are the sum of both fitted Gaussian components and are continuum subtracted. The empty East region in the H$_2$(1-0)Q(1) line is due to a lack of spectral coverage in that region.}
    \label{fig:line_maps}
\end{figure*}

We present the results of our spaxel-by-spaxel two component Gaussian fitting, as described in Section~\ref{sec:fitting}, for all detected H$_2$ lines in Figure~\ref{fig:line_maps}.
The Eastern side of the H$_2$(1-0)Q(1) map is empty as this region does not spectrally cover the entire emission line.

\bibliography{nifs_bib.bib}
\bibliographystyle{aasjournal}

\end{document}